# Magnetic properties of pseudomorphic epitaxial films of Pr$_{0.7}$Ca$_{0.3}$MnO$_3$ under different biaxial tensile stresses

A. Geddo Lehmann,[1] F. Congiu,[1] N. Lampis,[1] F. Miletto Granozio,[2] P. Perna,[2] M. Radovic,[2] and U. Scotti di Uccio[2]

[1]*Dipartimento di Fisica e CNISM, Università di Cagliari, S.P. Monserrato-Sestu, km 0.700, I-09042 Monserrato, Cagliari, Italy*
[2]*CNR-SPIN e Dip. Scienze Fisiche, Università di Napoli Federico II, Complesso Monte S. Angelo, Via Cinthia, I-80126 Napoli, Italy*

In order to analyze the effect of strain on the magnetic properties of narrow-band manganites, temperature- and field-dependent susceptibilities of about 8.5-nm-thick epitaxial Pr$_{0.7}$Ca$_{0.3}$MnO$_3$ films, respectively, grown on (001) and (110) SrTiO$_3$ substrates, have been compared. For ultrathin samples grown on (001) SrTiO$_3$, a bulklike cluster-glass magnetic behavior is found, indicative of the possible coexistence of antiferromagnetic and ferromagnetic phases. On the contrary, ultrathin films grown on (110) substrates show a robust ferromagnetism with a strong spontaneous magnetization of about 3.4$\mu_B$/Mn atom along the easy axis. On the base of high-resolution reciprocal space mapping analyses performed by x-ray diffraction, the different behaviors are discussed in terms of the crystallographic constraints imposed by the epitaxy of Pr$_{0.7}$Ca$_{0.3}$MnO$_3$ on SrTiO$_3$. We suggest that for growth on (110) SrTiO$_3$, the tensile strain on the film $c$ axis, lying within the substrate plane, favors the ferromagnetic phase, possibly by allowing a mixed occupancy and hybridization of both in-plane and out-of-plane $e_g$ orbitals. Our data allow to shed some physics of inhomogeneous states in manganites and on the nature of their ferromagnetic insulating state.

## I. INTRODUCTION

The microscopic nature of the ferromagnetic insulating (FMI) state setting in at low-doping levels in the phase diagrams of several manganites[1,2] represents a yet unsolved puzzle. While the physical interaction (i.e., the double exchange) that leads to the ferromagnetic metallic (FMM) state in manganites is well known,[3,4] the reason why an insulating state can persist in a ferromagnetic manganite is less evident. One possible interpretation assumes the coexistence of a FM metallic phase and a non-FM insulating phase. This scenario requires the non-FM insulating phase to cover a large majority of the sample, in order to rule out the formation of percolation paths of the minority conducting phase.[5] The second possibility is that the insulating nature is intrinsic to the FM phase and is due, rather than to double exchange, to a ferromagnetic superexchange interaction, substained by the formation of a special ordering of $e_g$ orbitals setting in above the Curie temperature $T_C$.[2,6,7]

Within this frame, the manganite Pr$_{1-x}$Ca$_x$MnO$_3$, a narrow-band hole-doped orthorhombic A$_{1-x}$A$'_x$MnO$_3$ system with insulating character throughout the whole $x=[0,1]$ range, deserves a special attention. The ferromagnetic phases of this system are considered to be intrinsically insulating. Bulk Pr$_{1-x}$Ca$_x$MnO$_3$ is reported to be a homogeneous FMI for the quite wide composition range $0 \leq x < 0.3$.[8] Studies on the $x=0.25$ composition showed the underlying orbital ordering to be of LaMnO$_3$ type, i.e., alternate ordering of $3x^2-r^2$ and $3y^2-r^2$ Mn 3$d$ orbitals in the ($a$-$b$) plane of the orthorhombic *Pbnm* cell associated to the A-type layered antiferromagnetism (AFM).[6] In such an orbitally ordered (OO) state, the spin exchange was found to be rather isotropic ($J_{ab} \cong J_c$), a feature which distinguishes the FMI phase from the highly anisotropic A-type AFM phase and that shows, conversely, an interesting resemblance with the orbitally disordered FMM state of medium and wide band manganites.

The structural distortion in the passage to the OO state, at the orbital-ordering temperature ($T_{OO}$), was found to be quite small which, together with the high spin-exchange isotropy, suggests that the electron distribution in the FMI phase is also more isotropic than in the layered AFM ones. A similar situation is realized in other FMI phases of manganites, like La$_{0.88}$Sr$_{1.2}$MnO$_3$, for which a staggered OO of $e_g$ orbitals $d(x^2-y^2)$ and $dz^2$, with an electronic 3$d$ cloud spread toward the $z$ direction, has been proposed.[7,9] For intermediate Ca doping ($0.3 \leq x < 0.5$), the FMI phase of Pr$_{1-x}$Ca$_x$MnO$_3$ coexists in with an AFM insulating one, and the system shows a spectacular magnetoresistance effect (colossal magnetoresistance), with a field-induced insulator to metal transition at which the resistance dramatically drops under a magnetic field of a few Tesla.

Within this work, we will address the properties of Pr$_{1-x}$Ca$_x$MnO$_3$ at its borderline ($x=0.3$) doping level. Pr$_{0.7}$Ca$_{0.3}$MnO$_3$ represents an outstanding example of the complexity of electron correlations in narrow-band manganites. According to,[10,11] the bulk Pr$_{0.7}$Ca$_{0.3}$MnO$_3$ separates, below room temperature, into two orthorhombic phases. The evolution of the peak widths in high-resolution neutron powder diffraction data implies that both phases are structurally well distinct from the room-temperature phase and that they are correlated on a mesoscopic length scale (50–200 nm). The first phase, showing a charge-ordering temperature $T_{CO}=220–250$ K and an antiferromagnetic transition at $T_N \approx 140$ K, is a phase with orbital ordering and strong charge localization. It is characterized by a relatively short $c_o$ lattice parameter [$c_o \approx 0.766$ nm with lattice spacing $d(002) < d(110)$] and is called CO phase in the following according to Ref. 11. The second phase, showing a FM transition at about 110 K,[12] presents a preferential orientation (i.e., larger than random) of the filled Mn 3$de_g$ orbitals with lobes perpendicular to the $a_o$-$b_o$ plane. It displays an intermediate degree of charge localization, a longer $c_o$ lattice parameter [$c_o \approx 0.768$ nm with $d(002) \approx d(110)$] and is called *reverse* OO (ROO) in the following according to Ref. 11. The overall magnetic response of such nanoscale-separated system, in low fields, was interpreted as resulting from an inhomogeneous mixture of ferromagnetic clusters in an antiferromagnetic matrix,[13] even though it shares phenomenological features with both spin glasses and superparamagnets. The average magnetic moment of about $2\mu_B$/Mn atom at 4 K indicates a FM volume fraction of about 50%. The application of magnetic field not only results in the melting of the CO phase as reported by Yoshizawa *et al.*[14] but also melts the FMI phase giving rise to a third FMM phase with a much smaller unit-cell volume.[11]

The orthorhombic insulating AFM and FM phases of $Pr_{0.7}Ca_{0.3}MnO_3$ both derive from a slight deformation of the parent (paramagnetic) cubic perovskite. As the magnetic and electronic properties of manganites are very sensitive to small structural distortions, they may be tuned by depositing epitaxial films under different biaxial stress conditions. In this paper we present results about films of $Pr_{0.7}Ca_{0.3}MnO_3$ grown on $SrTiO_3$ substrate with deposition cubic planes (110) and (001), which impose to the manganite different crystallographic deformations.[15] We report on the epitaxial stabilization of the FM phase and the concomitant suppression of the AFM one in $Pr_{0.7}Ca_{0.3}MnO_3$ films grown on (110) $SrTiO_3$, in contrast to the bulklike features retained in samples grown onto (001) $SrTiO_3$.

In a more general framework, the present investigation may be considered as an attempt to add a tile to the wide mosaic of the physics of inhomogeneous states in manganites, the ground state of which is indeed, in a broad region of parameter space, a nanoscale mixture of phases, with unavoidable disorder at phase boundaries. $Pr_{0.7}Ca_{0.3}MnO_3$ shows up as a case study of the role of such disorder, that determines glassy behavior also observed in our experiments and represents at the same time the best example of the extreme sensitivity of manganites to external (electrical, magnetic, and strain) fields, possibly supporting a description in terms of complexity and chaos. In this view, the great attention devoted to the use of the strain field as a tool to manipulate the macroscopic properties of manganites is easily understood. For instance, the effect of the substrate orientation on the charge-ordering transition of the narrow-bands manganites $Bi_{0.4}Ca_{0.6}MnO_3$ and $Nd_{0.5}Sr_{0.5}MnO_3$ has been deeply investigated.[16,17] In the present work, we resorted to measurements of magnetization as a function of temperature on ultrathin $Pr_{0.7}Ca_{0.3}MnO_3$ films, focusing our attention on the ferromagnetic state and on its dependence on the applied stress.

## II. EXPERIMENTAL

The epitaxial $Pr_{0.7}Ca_{0.3}MnO_3$ films were deposited by pulsed laser deposition in a multichamber system devoted to the deposition of oxide epitaxial thin film and their *in situ* surface characterization. Further details are reported in Ref. 18. The growth process was assisted by high-pressure-reflected high-energy electron diffraction (RHEED), which demonstrated that the two-dimensional growth is preserved on both (001) and (110) $SrTiO_3$ until the end of the deposition of films as thick as 10 nm.[19] The samples investigated in this work are all thinner than 10 nm. In the following, our two samples grown onto the two different substrate orientations and analyzed in this work will be called F_001 and F_110. The crystallographic characterization was performed by high-resolution x-ray diffraction (XRD), through the analysis of $\omega$ scans, longitudinal $\omega$-$2\theta$ scans and reciprocal space maps (RSMs), using a four circles diffractometer (Bruker D8 Discover) with Cu anode, equipped with a Goebel mirror and a 2-bounces monochromator. The dc (i.e., static field) magnetization measurements were performed by a Quantum Design MPMS5 XL5 superconducting quantum interference device magnetometer with a sensitivity of $10^{-8}$ emu, equipped with a superconducting magnet producing fields up to 50 kOe and calibrated using a Pd standard. The residual field during the zero-field cooling was less than $5 \times 10^{-2}$ Oe. The temperature dependence of magnetization $M(T)$ was measured through a standard zero-field-cooled-field-cooled (ZFC-FC) procedure: (i) cooling the sample from high temperature in zero field down to the lowest temperature, (ii) switching on the measuring field $H$ and recording the magnetization $M_{ZFC}(T)$ on warming, and (iii) cooling the sample in the same field and recording $M_{FC}(T)$.

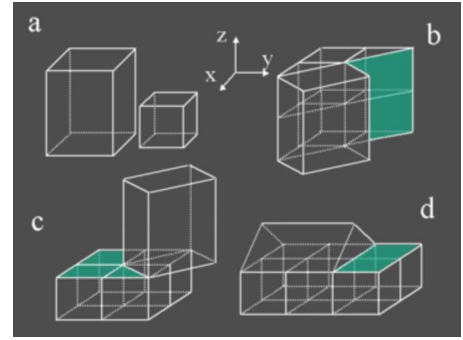

FIG. 1. (Color online) Epitaxy of $Pr_{0.7}Ca_{0.3}MnO_3$ on $SrTiO_3$. (a) orthorhombic *Pbnm* and cubic $Pm\bar{3}m$ cells of $Pr_{0.7}Ca_{0.3}MnO_3$ and $SrTiO_3$; (b) epitaxy of F_110; (c) epitaxy of F_001 in the *V1* variant; (d) epitaxy of F_001 in the *V2* variant. The shaded areas indicate the substrate surface.

## III. STRUCTURAL CHARACTERIZATION

In Fig. 1 we depict the epitaxy of $Pr_{0.7}Ca_{0.3}MnO_3$ on (001) $SrTiO_3$ and (110) $SrTiO_3$, as it results from our x-ray analysis. The orthorhombic *Pbnm* cell of $Pr_{0.7}Ca_{0.3}MnO_3$ and the cubic cell of $SrTiO_3$ ($a_s$=0.3905 nm) are sketched in Fig. 1(a). As a result of the $MnO_6$ octahedra deformation and rotation, the cell of $Pr_{0.7}Ca_{0.3}MnO_3$ may be seen as a superstructure containing four pseudocubic perovskitic cells. The orthorhombic lattice parameters $a$, $b$, and $c$ are related to the lattice spacing $a_c$ of the parent cubic phase with $Pm\bar{3}m$ structure by $a \cong \sqrt{2}a_c$, $b \cong \sqrt{2}a_c$, and $c \cong 2a_c$. The values of the room-temperature lattice parameters of bulk $Pr_{0.7}Ca_{0.3}MnO_3$ are:[20] $a_o$=0.5426 nm, $b_o$=0.5478 nm, and $c_o$=0.7679 nm (with the subscript "o" used henceforth to indicate the bulk).

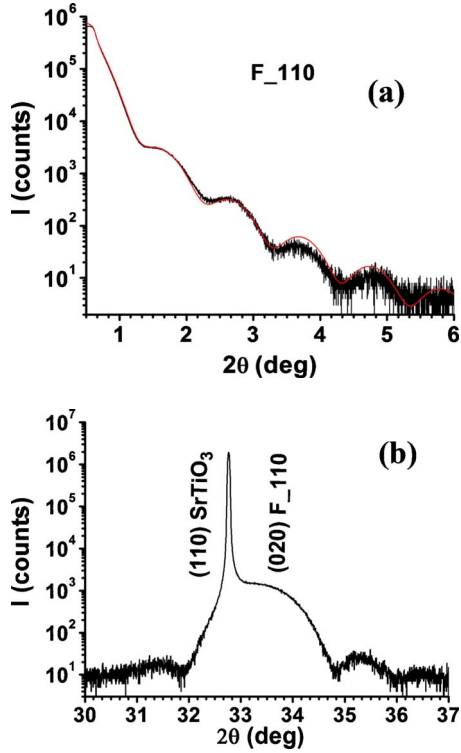

FIG. 2. (Color online) F_110: (a) x-ray reflectivity curve, (b) symmetric $\omega$-$2\theta$ scan around the (110) SrTiO$_3$ peak [film rocking curve full width at half maximum (FWHM)=0.02°].

The quantitative results of our XRD characterization (Figs. 2–4) are summarized in Table I, in which we report, for both samples, the film-to-substrate relative orientation (as discussed in the following two sections), the strained film lattice parameters and $e_1$, $e_2$, and $e_3$ film misfit strains, defined in terms of lattice parameters as the deformation needed to transform the actual film (no subscript) lattice parameters into the corresponding *Pbnm* bulk values, i.e., $e_i = \frac{a_i}{a_{oi}}/-1$, with $a_1 = a$, $a_2 = b$, and $a_3 = c$. Compressive misfit strain gives $e_i < 0$ while tensile misfit strain corresponds to $e_i > 0$. The film thickness was determined from x-ray reflectivity measurements [that we show in Fig. 2(a) in the case of F_110], resulting in (8.4±0.2) nm and (8.6±0.2) nm for F_110 and F_001, respectively.

### A. Growth of the F_110 film

The epitaxial growth of a perovskite with *Pbnm* structure on a cubic perovskite substrate is such to make the above-mentioned parent cubic phase align "cube on cube" on the substrate. According to the details of the octahedra tilting, the *Pbnm* unit cell of Pr$_{0.7}$Ca$_{0.3}$MnO$_3$ can assume on (110) SrTiO$_3$ six different epitaxial configurations. Four of them have the (112) Pr$_{0.7}$Ca$_{0.3}$MnO$_3$ planes aligned with (110) SrTiO$_3$. The other two are the "*a*-axis" and the "*b*-axis" orientations.

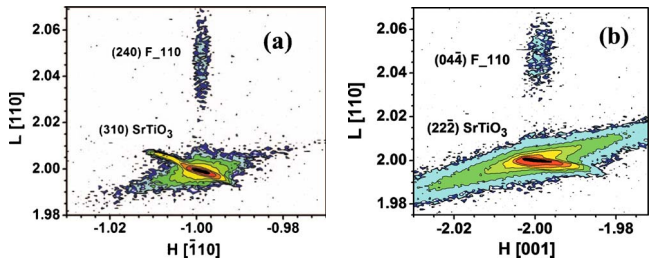

FIG. 3. (Color online) F_110: (a) RSM around the (310) SrTiO$_3$ peak, (b) RSM around the (222) SrTiO$_3$ peak.

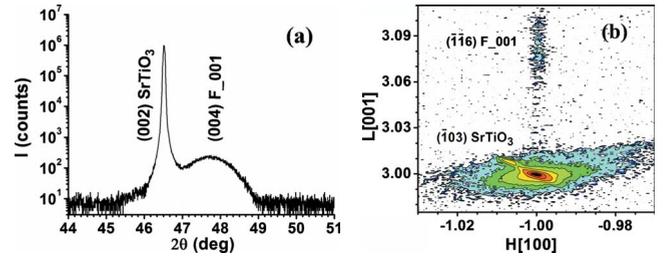

FIG. 4. (Color online) F_001: (a) symmetric $\omega$-$2\theta$ scan around the (002) SrTiO$_3$ peak (film rocking curve FWHM=0.02°), (b) RSM around the ($\bar{1}$03) SrTiO$_3$ peak.

Identifying the orientation of a strained Pr$_{0.7}$Ca$_{0.3}$MnO$_3$ epitaxial films is not straightforward, as already noticed by other authors on other orthorhombic *Pbnm* manganites,[21] since the directions can neither be directly determined by the oxygen octahedra tilting, that is out of the reach of standard XRD film measurements, nor by the values of the measured lattice parameters, that are modified by strain. A first qualitative argument based on surface-energy minimization[22] suggests to rule out the (112)-oriented configurations, because of the high surface energy of this high index plane. A simple data analysis based on the measured interplane distances, furthermore, demonstrates that the *a*-axis orientation allows to fit the experimental data with the overall minimum strain. Since the interplay of strain and orientation is well established in this class of ferroelastic systems, the sample is henceforth assumed to be *a*-axis oriented. In F_110, two inequivalent in-plane orthogonal directions exist, parallel to the substrate edges [001]$_s$ and [$\bar{1}$10]$_s$ (here and in the following, the subscript "s" is adopted to indicate the substrate). According to the previous considerations, the axes orienta-tion of F_110 with respect to the substrate [Fig. 1(b)] is $a\|[110]_s$, $b\|[\bar{1}10]_s$, and $c\|[001]_s$. The length of the out-of-plane $a$ axis is determined by the $\omega$-$2\theta$ scan [Fig. 2(b)]. To obtain the two orthorhombic in-plane lattice parameters $b$ and $c$, RSMs around pseudocubic (222) and (130) substrate asymmetric reflections were collected. Results are displayed in Figs. 3(a) and 3(b), in which orthorhombic Miller indexes are used for the film. Both $b$ and $c$ axes were found to be matched to the substrate. The details of the lattice parameters and unit-cell deformation are reported in Table I.

TABLE I. Relative orientation between F_001 and F_110 and substrate; strained film lattice parameters $a$, $b$, and $c$; $e_1$, $e_2$, and $e_3$ film misfit strains (see text for definition). (a) under the assumption $[110] \perp [\bar{1}10]$.

| | F_110 | F_001 V1 | F_001 V2 |
|---|---|---|---|
| Out of plane | $a \parallel [110]_s$ | $c \parallel [001]_s$ | $[110]_F \parallel [001]_s$ |
| In plane | $b \parallel [\bar{1}10]_s$ | $a \parallel [110]_s$ | $[\bar{1}10]_F \parallel [010]_s$ |
| | $c \parallel [001]_s$ | $b \parallel [\bar{1}10]_s$ | $c \parallel [100]_s$ |
| $a$ (nm) | 0.5410(1) | 0.5522(1) | 0.5466(1), $\gamma = 91.5°$ (a) |
| $b$ (nm) | 0.5522(1) | 0.5522(1) | 0.5466(1) |
| $c$ (nm) | 0.7810(1) | 0.7620(1) | 0.7810(1) |
| $e_1$ | −0.003 | +0.018 | +0.007 |
| $e_2$ | +0.008 | +0.008 | −0.002 |
| $e_3$ | +0.017 | −0.008 | +0.017 |

### B. Growth of the F_001 film

For the F_001 film, by applying the above-mentioned concept of cube-on-cube epitaxy, we obtained again six possible configurations, that can be grouped in two principal epitaxial variants, named *V1* and *V2* (Refs. 23 and 24) in the following. In the *V1* phase, the film grows with its (001) plane parallel to the (001) surface of SrTiO$_3$ [Fig. 1(c)] with the in-plane $a$ and $b$ axes oriented according to $a\|[110]_s$ and $b\|[\bar{1}10]_s$. Exchange between $a$ and $b$ gives two equivalent domains. In the *V2* variant, the film aligns its (110) plane parallel to the (001) surface of SrTiO$_3$, and the in-plane axes have orientations $c\|[010]_s$ and $[\bar{1}10]_f\|[010]_s$ (the subscript "f" standing here for the film direction), as displayed in Fig. 1(d). It is easy to realize that *V2* has four equivalent domains. In a fully strained Pr$_{0.7}$Ca$_{0.3}$MnO$_3$ film, the orthorhombic $c$ axis would be compressed in the *V1* variant. On the contrary, it would be elongated in *V2*. The orthorhombic in-plane anisotropy would be metrically quenched in the *V1* variant. The *V2* variant would present a monoclinic distortion, resulting from the epitaxial deformation of the rectangular $a$-$b$ basal plane of the bulk *Pbnm* cell, with an angle $\gamma$ between $a$ and $b$ slightly greater than 90° [Fig. 1(d)]. We note also that *V2* can match the $(1\bar{1}0)$SrTiO$_3$ interplanar distance in different ways, by changing both $a$ and $b$ lattice parameters and the angle $\gamma$. Once again, the actual fraction of *V1* and *V2* variants cannot be unequivocally determined from our x-ray patterns. In the analysis of our x-ray data, we first observe that F_001 [the out-of-plane interplanar distance of which is determined from the $\omega$-$2\theta$ scan of Fig. 4(a)] is fully strained, as shown by the RSM of Fig. 4(b) (here again, orthorhombic indexes are used for the film). We separately reported in Table I the lattice parameters calculated in the case of two distinct assumptions: that either (a) the *V1* variant or (b) the *V2* variant is realized. In Table I, we assumed $[110]_f \perp [\bar{1}10]_f$ as in Ref. 25, meaning that both $a$ and $b$ and $\gamma$ have been modified by strain, a choice compatible within the error with our experimental results. We do not try to evaluate the shear strain for *V2*.

## IV. MAGNETIC CHARACTERIZATION

### A. Magnetic characterization of F_001

The ZFC-FC magnetization curves ($M_{ZFC}/M_{FC}$) of F_001 as a function of temperature are shown in Fig. 5(a), for fields in the range $25 < H < 5200$ Oe, applied parallel to $[100]_s$. A marked increase in the magnetic moment, that we interpret as a ferromagnetic transition, occurs at a Curie temperature $T_C$ of about 124 K, estimated from the curve recorded during cooling in $H=0$ Oe. On cooling, a ZFC-FC irreversibility develops at $T_{irr}$, with a well-defined maximum in $M_{ZFC}$ at $T_{max} < T_{irr}$. The $M(T)$ plots at $H=5200$ Oe, reproduced in enlarged vertical scale in Fig. 5(b), reveal other features, as indicated by the arrows. The marked peak at $T_{CO}=230$ K is the signature of the charge-ordering transition. The transition temperature is very close to the bulk value and coincides with the one determined by resistivity measurements[19] on samples of the same batch. Concerning the weak feature visible at 190 K, we can refer to what happens in the bulk, in which antiferromagnetism sets in between the charge-ordering temperature and the Curie point ($T_C < T_N < T_{CO}$). Analogously, 190 K might represent the Néel point of F_001, with the magnetization cusp masked, on the low-temperature side, by the tail of the ferromagnetic response, which is present under the measuring field of 5200 Oe. Finally, Fig. 5(b) shows a strong ZFC-FC irreversibility for temperatures between $T_{CO}$ and about 380 K. In this temperature range, a crossover from a charge localized to the charge-ordered state occurs in the bulk, with the formation at $T_p \approx 400$ K of small polarons of Jahn-Teller (JT) origin, with associated short-range ferromagnetic correlations.[26] These effects stem from the strong spin-lattice coupling in this colossal magnetoresistance manganite.

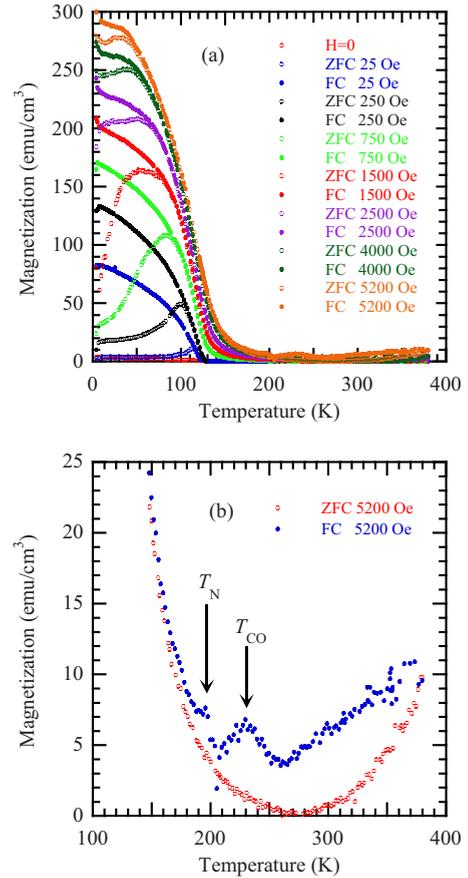

FIG. 5. (Color online) (a) ZFC-FC magnetization ($H$ in plane) of the 8.4-nm-thick F_001 sample, (b) magnification of the ZFC-FC magnetization at $H=5200$ Oe.

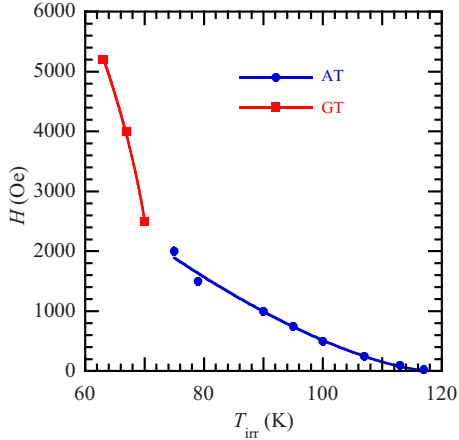

FIG. 6. (Color online) Irreversibility temperature $T_{\text{irr}}$ vs $H$. The fit to the AT model is the blue line on dots; the fit to the GT model is the red line on squares.

The magnetization values measured at 4 K for different applied fields [Fig. 5(a)] are comparable to those reported for the bulk, in which $M(H)$ at 4 K saturates for $H \approx 5000$ Oe.[13] The magnetic moment of F_001 at 4 K and $H=5200$ Oe corresponds to about $1.7 \mu_B$/Mn site. This value is considerably lower than $3.7 \mu_B$/Mn site expected for a single FM phase and is close to the one reported for the bulk.[13] As for the bulk, the reduced magnetic moment per Mn site observed in F_001 can be related to the presence of a competing AFM second phase. This is also suggested by the observed type of branching in $M_{\text{ZFC}}/M_{\text{FC}}$ and by the presence of the peak in $M_{\text{ZFC}}$, which are common findings in a variety of magnetically disordered systems characterized by competing FM and AFM interactions at different length scales, ranging from homogeneous spin glasses to cluster glasses. An important point to be noted here is that $M_{\text{FC}}$ of F_001 strongly increases below the irreversibility temperature $T_{\text{irr}}$. This is a typical feature of various cluster glass systems,[27,28] where FM interactions within the clusters tend to align the intracluster moments in the direction of the applied field, resulting in increased magnetization and in a large difference $\Delta M = M_{\text{FC}} - M_{\text{ZFC}}$. Such behavior differs from the case of homogeneous spin glasses, where the spins are paramagnetic above $T_{\text{irr}}$ (the magnetic susceptibility showing therefore either a Curie or a Curie-Weiss temperature dependence above $T_{\text{irr}}$), and $M_{\text{FC}}$ reaches a plateau below $T_{\text{irr}}$.[29] Such cluster glassy behavior resembles that of bulk $Pr_{0.7}Ca_{0.3}MnO_3$, the magnetism of which has been described in terms of ferromagnetic clusters embedded in a AFM, charge-ordered matrix.[13]

The field dependence of $T_{\text{irr}}$ (Fig. 6) provides further insight on the magnetic nature of the cluster system. Below 2000 Oe, the data lie on a de Almeida-Thouless (AT) (Ref. 30) irreversibility line $H = A_{\text{AT}} [1 - \frac{T_{\text{irr}}(H)}{T_{\text{irr}}(0)}]^{3/2}$; at higher fields, $T_{\text{irr}}$ shows a steep increase. Tentatively, a Gabay-Toulouse (GT) (Ref. 31) line $H = A_{\text{GT}} [1 - \frac{T_{\text{irr}}(H)}{T_{\text{irr}}(0)}]^{1/2}$ is suggested as a fit in this region. The faith in the fitting cannot be pushed too far, but the phenomenology points to the existence of a crossover, possibly related to the anisotropy of the system, from a behavior qualitatively consistent with the Ising model, to a behavior better described by the Heisenberg model.[32,33] It may be assumed that the observed $M_{\text{ZFC}}(T)$ maximum in F_001 is governed by a local anisotropy field acting on the spins inside each FM cluster, as proposed for other perovskites with cluster-glass or re-entrant spin-glass behavior.[34–36] In this view, the magnetic moments of the spins would be frozen in directions energetically favored by their local anisotropy or by the external field, respectively, when the system is cooled down from high temperature in a zero or nonzero field, respectively. The change in the exponent $n$ from the AT to the GT model is related to the different sensitivity of the local anisotropy to temperature. This mechanism, originated from the growing size of ferromagnetic clusters, was invoked to explain the properties of $Pr_{0.7}Ca_{0.3}MnO_3$ bulk[13] and it also seems appropriate here.

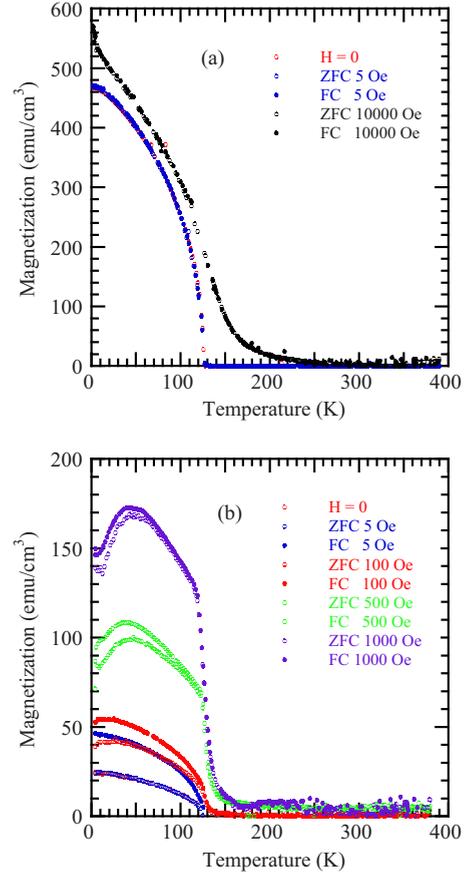

FIG. 7. (Color online) ZFC-FC magnetization of F_110, with $H$ applied along the two principal in-plane directions parallel to the substrate edges. (a) easy-axis magnetization, (b) harder axis magnetization.

### B. Magnetic characterization of F_110

The ZFC-FC magnetizations of F_110, recorded with $H$ applied along the two in-plane directions parallel to the substrate sides, are reported in Fig. 7. The Curie temperature, estimated from the lowest field curves, is $T_C = 127$ K, higher than that reported for the bulk of 110 K. The peak at $T_{\text{CO}} \approx 230$ K appearing in $M_{\text{FC}}(T)$ recorded at $H=1000$ Oe along the hard direction, shown enlarged in Fig. 8, can be attributed to a charge localization, while it is very difficult to assert the establishment of a precursor lattice process of Jahn-Teller origin above $T_{CO}$, even if a certain increase in the fluctuation of magnetization signal can be observed in the high-temperature range.

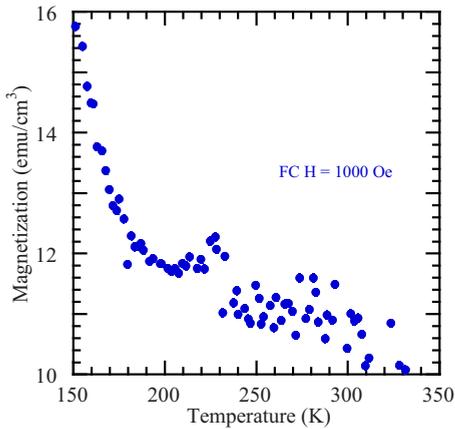

FIG. 8. (Color online) Magnification of the field-cooled magnetization $M_{FC}$ of F_110 along the harder axis at $H=1000$ Oe.

The measurements along the two in-plane directions reveal a strong anisotropy, which can be phenomenologically described as driven by magnetostriction.[37] For $H$ along the easy axis [Fig. 7(a)], the branching of ZFC-FC measurements is absent. The spontaneous ($H=0$ Oe) magnetization $M(T)$ at 4 K exceeds by over one order of magnitude the case of F_001 sample. $M(H=0$ Oe) amounts to over 80% of the saturated $M_{FC}$ value measured at an applied field $H=10^4$ Oe, that is $3.4\mu_B$/Mn site. We note that this saturation value is considerably larger than in the bulk case, that is $\sim 2\mu_B$ (Refs. 10 and 14) and in the F_001 film, that is $\sim 1.7\mu_B$. This is consistent with the idea that most of Mn atoms in F_110 participate to the FM alignment. The previous results show that the bulklike, cluster-glass behavior, retained in F_001, is replaced in F_110 by a robust ferromagnetism.

When $H$ is applied along the orthogonal in-plane direction [Fig. 7(b)], both $M_{FC}$ and $M_{ZFC}$ deviate significantly from a well-behaving ferromagnetic $M(T)$ order parameter, showing, for increasing applied fields, a reduced growth rate of the magnetic moment from just below the Curie point down to about 50 K, temperature at which broad maxima appear in both ZFC and FC branches, which run almost parallel in the whole measured temperature range. This anomalous $M(T)$ behavior deserves a specific comment. We observe that $M(T)$ maxima take place not far from 60 K, a temperature recurrent in the Pr-Ca-Mn-O system and that marks (in bulk $Pr_{0.7}Ca_{0.3}MnO_3$) the onset of several different interesting phenomena. In particular, 60 K is the temperature below which ordering of $Pr^{3+}$ 4f moments has been reported,[10] and below which the magnetic-field-induced insulator to metal transition is irreversible.[38–40] The strong magnetization anisotropy that we observe in F_110 for the two orthogonal directions of $H$ is, on the other hand, reminiscent of the anisotropic behavior which is typical of the $Pr^{3+}$ Van-Vleck paramagnetic susceptibility.[41,42] We propose therefore that the observed features in $M(T)$ between $T_C$ and 50 K may be connected to the $Pr^{3+}$ species and to a possible interaction between the rare-earth and the manganese sublattices, switched on just below the ferromagnetic Mn transition.

## V. DISCUSSION

The *in situ* RHEED and *ex situ* XRD structural characterization of F_001 and F_110 indicate complete in-plane matching to the substrates, with lattice parameters quite different from the bulk ones, which suggests that the observed properties are due to the strain state. On the other hand, the transition temperatures of both samples (very close to the bulk ones or even enhanced) do not indicate finite-size scaling. This guarantees that the observed ferromagnetism is not due to uncompensated moment of antiferromagnetic phases, a feature typical of ultrathin AFM layers and most evident in those AFM phases composed by alternate FM planes.[43] We note that the depressed ferromagnetism in F_001 with respect to F_110 could, in principle, be due to the formation of a nonmagnetic dead layer. Actually, a nonmagnetic layer few nanometer thick at the substrate/film interface is often observed in manganite films, depending on the method and growth conditions.[44] The origin of the dead layer is still debated and different mechanisms of formation have been invoked. The role of the substrate as a source of disorder within the first layers of the film, underlined in Ref. 45, can be excluded for our samples, since we used atomically flat and single-terminated (001) substrate surface.[18] A dead layer due to cation segregation during strain accommodation, as reported by Ref. 46 for $La_{0.7}Ca_{0.3}MnO_3$ on $SrTiO_3$ (001) grown by RF sputtering, can be also ruled out because cation segregation requires migration of species with different size, while $Pr^{3+}$ and $Ca^{2+}$ have almost equal ionic radii (0.179 nm and 0.180 nm, respectively), so much so that $Pr_{1-x}Ca_xMnO_3$ is considered a case study of manganite with negligible quenched A-size disorder.[47] But it is the observed cluster-glass magnetic behavior itself, that makes a strong case toward a scenario in which FM and AFM small regions coexist in F_001 on nanometric length scale, rather than a description of the film as separated into a nonmagnetic layer and a ferromagnetic one or with gradual change in the Ca/Pr ratio across the film thickness. Given the previous considerations, we ascribe the differences between the physical properties of F_001 and F_110 samples only to their different strain state due to the different crystallographic orientation of the substrate.

The key role played by structural deformations in the physics of manganites is well assessed.[48] The physical properties of these materials depend on the overlap between the manganese $d$ orbitals and oxygen $p$ orbitals, which are closely related to the Mn-O-Mn bond angles and the Mn-O distances. As the unit cell of the thin film is modified with respect to the bulk material, the Mn-O distances and Mn-O-Mn angles are altered, inducing variations in the electronic and magnetic properties. In $Pr_{0.7}Ca_{0.3}MnO_3$, as in many narrow-band manganites, the distortion from the parent cubic symmetry $Pm\bar{3}m$ is described by a combination of tilting and deformation of the $MnO_6$ octahedra.

The tilting of the octahedra is related to the tolerance factor $t=(r_A+r_O)/\sqrt{2}(r_{Mn}+r_O)$, expressed in terms of atomic radii, and causes the transition from the parent cubic $Pm\bar{3}m$ structure to an orthorhombic structure (space group *Pbnm*). Such transition can be described by two order parameters, i.e., the rotation angle $\vartheta$ (commonly indicated as tilt angle) of the $MnO_6$ octahedra around the orthorhombic $b_o$ axis and their rotation angle $\phi$ around $c_o$, as shown in Fig. 9.

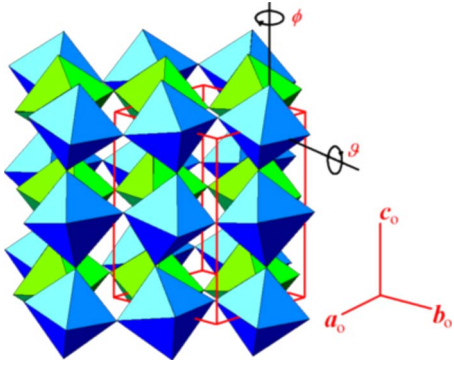

FIG. 9. (Color online) Octahedra tilting in the *Pbnm* space group. The tilt angle $\vartheta$ (rotation around $b_o$) and the rotation angle $\phi$ around $c_o$ are indicated.

Tilt causes the buckling of the Mn-O-Mn bonds, which, in turn, reduces the hopping integral $t_{ij}$ of the $e_g$ electrons between neighboring sites, thus weakening double-exchange ferromagnetism. At room temperature and ambient pressure, the bond angles of bulk $Pr_{0.7}Ca_{0.3}MnO_3$ are $\theta_1 = 180° - 2\vartheta$ =Mn-O1-Mn=156.5° and $\theta_2 = 180° - 2\phi$=Mn-O2-Mn =156.4°, O1 denoting the two apical oxygen atoms and O2 the four planar ones.

The deformations of the $MnO_6$ units are related to the Janh-Teller effects, induced by the $Mn^{3+}$ ion and acquire a cooperative character at the cooperative Jahn-Teller ("orbital ordering") transition. The JT distortion also competes with double exchange: at $Mn^{3+}$ sites, the two JT modes $Q_3$ and $Q_2$, of tetragonal and orthorhombic symmetries, respectively,[49] stabilize local structural distortions that, for strong electron-phonon coupling, may trap the electrons in particular orbitals, favoring their localization. The bond distances of bulk $Pr_{0.7}Ca_{0.3}MnO_3$ at room temperature, defining the internal JT distortion of the $MnO_6$ groups, are $d(\text{Mn-O1})=0.1964$ nm, $d(\text{Mn-O2})=0.1967$ nm, and 0.1976 nm.[50] The degree of anisotropy of the Jahn-Teller distortion can be evaluated by the ratio $Q_3/Q_2$, expressed in terms of the short ($s$), medium ($m$), and long ($l$) Mn-O distances as $\tan\Phi=(2m-l-s)/\pm\sqrt{3}(l-s)$.[49] The highest anisotropy, corresponding to pure $Q_3$ distortion, gives $\Phi=30°$. For bulk $Pr_{0.7}Ca_{0.3}MnO_3$, $\Phi$ is about 21°, denoting a relevant contribution of the tetragonal mode. The structural distortion and the Jahn-Teller anisotropy in bulk $Pr_{0.7}Ca_{0.3}MnO_3$ are so high that the double-exchange metallic state is not stable at any temperature.

Unfortunately, no information about the way in which angles variation and bond distances modification contribute to accommodate the strain in our epitaxial $Pr_{0.7}Ca_{0.3}MnO_3$ films is available. However, it is known that, under hydrostatic pressure, $d(\text{Mn-O1})$ is modified quite easily while the two distances $d(\text{Mn-O2})$ remain almost unchanged (with bulk linear compressibilities $k_c=0.0032$ $GPa^{-1}$ and $k_a$, $k_b$ =0.0015 $GPa^{-1}$).[51] It is also known that pressure can modify the mean bond angle $\theta=(\theta_1+2\theta_2)/3$, but the way $\theta_1$ and $\theta_2$ change separately is not known.

Concerning the robust ferromagnetic phase observed in F_110, the most interesting result of our investigation, the key question arises whether it is a bulklike insulating phases, with (superexchange) ferromagnetism sustained by orbital ordering, or if its origin can be found in a nonbulklike double-exchange mechanism. If the observed FM phases is bulklike, the striking difference in the magnetic behavior of F_001 and F_110 could be interpreted as due to the different mesh and interconnection between FMI clusters embedded in the AFM insulating OO matrix, with the strain imposed by the (110) substrate promoting the growth of the FMI cluster almost over the entire volume. As mentioned in the introduction, the analysis of bulk, phase-separated $Pr_{0.7}Ca_{0.3}MnO_3$ singled out the presence of a FM ROO phase with a longer $c_o$ axis and an AFM CO phase with a shorter $c_o$ axis.[11] Since the two phases are almost degenerate in energy, and since their relative amount in the bulk is assumed to be determined, among other factors, by the reciprocal strain, it can be expected that for epitaxial thin films the boundary conditions on a rigid substrate might play a role in promoting one of the phases with respect to the other. In the case of the F_110 film, we observe that the $c$ axis of $Pr_{0.7}Ca_{0.3}MnO_3$ is under tensile strain (see Table I), possibly favoring the FM ROO phase. We note, also, that a bulklike OO would be allowed by symmetry in F_110, because the constraint imposed by $SrTiO_3$ (110) maintains an orthorhombic cell, although with nonbulk lattice parameters. It is known that, in Jahn-Teller systems, the orthorhombic component $Q_2$ of the crystal field stabilizes hybrid orbitals expressed as a superposition of $e_g$ orbitals $d(x^2-y^2)$ and $dz^2$, that can order in a staggered (antiferrodistortive) pattern compatible with each Mn ions having all six first neighbors coupled by ferromagnetic superexchange integrals, as reported for FMI $Pr_{0.75}Ca_{0.25}MnO_3$ and $La_{0.88}Sr_{0.12}MnO_3$.[6,7] In the alternative scenario, the ferromagnetism we measured in our ultrathin F_110 films could be due to double exchange. In this case, no orbital ordering would set in. Actually, we do not observe in F_110 the feature in $M(T)$ above $T_{CO}$ associated to the precursor Jahn-Teller polaron formation. On the other hand, the hypothesis of a strain-induced suppression of the cooperative Jahn-Teller effect in F_110 appears to be reasonable on the base of crystallographic considerations. We recall that bulk $Pr_{0.7}Ca_{0.3}MnO_3$ has apically compressed $MnO_6$ octahedra, which is the reason of the stabilization of layered AFM phases. Taking into account the fact that $d(\text{Mn-O1})$ can be modified more easily than $d(\text{Mn-O2})$,[51] the strain imposed by (110) $SrTiO_3$ on the Mn-O distances, which is tensile on $d(\text{Mn-O1})$, appears to be well suited to contrast the apical compression of $MnO_6$ units and, by consequence, to weaken the Jahn-Teller electron localization which takes place in the bulk. The strain effect on bond angles in F_110 and the consequences on the $e_g$ bandwidth $W$ must also be considered. For a double-exchange system, the width of the tight-binding band for itinerant $e_g$ electrons, mediated on the three directions, is $W_\sigma = \varepsilon_\sigma \lambda_\sigma^2 \cos(\pi - \theta) \langle \cos(\Theta_{ij}/2) \rangle$, in which $\varepsilon_\sigma$ is the one electron energy, $\lambda_\sigma$ is the overlap integral, and $\Theta_{ij}$ is the angle formed by localized $t_{2g}$ core spins.[52] If we suppose— for a moment—that in F_110 $d$(Mn-O1) maintains the bulk value, by using the measured lattice parameter $c$ =0.7810 nm we determine a Mn-O1-Mn bond angle $\theta_1$ of 167.6°, much greater than the bulk one. Even if part of the tensile strain along $c$ were accommodated by an increase in $d$(Mn-O1), the increase in $\theta_1$ would still be impressive: for instance, $\theta_1$ would be as great as 164.7° if $d$(Mn-O1) reached a value comparable with $d$(Mn-O2) (i.e., about 0.197 nm), corresponding to almost regular octahedra. With such an opened Mn-O1-Mn angle, the extra $e_g$ electrons are likely to be easily delocalized on Mn pairs along $[001]_f$,[53] leading to the formation of manganese dimers coupled ferromagnetically by double exchange (Zener polarons).[54] This would open a double-exchange channel favoring the FM phase against the electron-localized AFM matrix. Any further consideration about the effect of bond angles on the electronic and magnetic properties requires, however, at least a rough estimate of $\theta_2$, which determines the bandwidth within the (001) plane. Under the approximation of regular octahedra[55] we obtain for F_110 $\theta_2$=154.5°, leading to an average bond angle $\theta$ appearing in the expression of the mean bandwidth of 157.9°. We note that this value is very close to the critical average Mn-O-Mn angle ($\theta_c \approx 158°$) that marks in manganites, at hole-doping level $x$=0.3, a crossover from high to low ferromagnetic moment per Mn site. In other words, in a phase-separation scenario, $\theta_c$ marks the passage from a homogeneous FM phases to a phase-separated (AFM+FM) or canted states.[53] The observed robust, almost fully polarized ferromagnetic behavior of F_110 is in quite good agreement with this semiempirical prediction.

Finally, we note that a correlation was found in manganites between the average Mn-O-Mn angle $\theta$ and the Jahn-Teller anisotropy ratio $Q_3/Q_2$,[53] according to which compounds with $\theta$ higher than $\theta_c$ have at the same time low MnO$_6$ anisotropy and inhibited tendency toward phase separation. If this also holds in the case of thin films, the strain imposed by SrTiO$_3$(110) to Pr$_{0.7}$Ca$_{0.3}$MnO$_3$ is playing against the tendency to separate two phases and helps to stabilize a single FM phase.

Analogous qualitative considerations on the effect of epitaxial strain can apply to the F_001 sample. The overall magnetic behavior is strongly reminiscent of bulk Pr$_{0.7}$Ca$_{0.3}$MnO$_3$, thus suggesting that the previously described AFM CO and a FM ROO phase might also coexist in our sample. Simple geometrical considerations show that the two phases might minimize their strain in the film by adopting different epitaxial variants. The *V*1 orientation might better accommodate the CO phase, due to the tensile strain in the $a$-$b$ plane while the *V*2 orientation might better accommodate the ROO phase, having a longer $c$ axis. Moreover, because of the equality of the $c$ parameters, the structural considerations made about the possible electronic band widening along $k_z$ in F_110 apply also to *V*2, suggesting that a double-exchange channel might be triggered in F_001, too.

## VI. SUMMARY

We studied ultrathin films of Pr$_{0.7}$Ca$_{0.3}$MnO$_3$ on SrTiO$_3$ with different crystallographic orientation, with the aim of characterizing the relation between the magnetic properties and the strain imposed by epitaxy. The XRD measurements confirm the structural quality of the samples and prove the pseudomorphic nature of the films. Magnetization $M(T)$ measurements show that films as thin as $\sim$9 nm have values of the transition temperatures to ordered states similar or even slightly enhanced with respect to the bulk. The ZFC-FC magnetization curves unveil a deep difference between the film deposited on (001) SrTiO$_3$ (F_001) and the one deposited on (110) SrTiO$_3$ (F_110), confirming the high sensitivity of narrow-band manganites to strain. We showed that the biaxial tensile stress with in-plane square symmetry does not change in a significant way the magnetic properties with respect to bulk, presumably leading to coexisting FM and AFM phases in F_001. In contrast, the biaxial tensile stress with in-plane rectangular symmetry promotes a robust FM state in F_110, showing a strong in-plane anisotropy. The saturated magnetization ($M$=3.4$\mu_B$) and an overall magnetic behavior vs temperature that is rather reminiscent of standard double-exchange magnetism in medium band manganites than of bulk Pr$_{0.7}$Ca$_{0.3}$MnO$_3$. We discussed our results with references to the structural deformation imposed by the epitaxy in terms of modified Jahn-Teller distortion of the MnO$_6$ units and of the Mn-O-Mn bond angles.

[1] M. Imada, A. Fujimori, and Y. Tokura, Rev. Mod. Phys. **70**, 1039 (1998).

[2] J. Geck, P. Wochner, S. Kiele, R. Klingeler, A. Revcolevschi, M. v Zimmermann, B. Büchner, and P. Reutler, New J. Phys. **6**, 152 (2004).

[3] P.-G. de Gennes, Phys. Rev. **118**, 141 (1960).

[4] E. Dagotto, Phys. Rep. **344**, 1 (2001).

[5] M. Uehara, S. Mori, C. H. Chen, and S.-W. Cheong, Nature (London) **399**, 560 (1999).

[6] R. Kajimoto, H. Mochizuki, and H. Yoshizawa, Physica B **329-333**, 738 (2003).

[7] Y. Endoh, K. Hirota, S. Ishihara, S. Okamoto, Y. Murakami, A. Nishizawa, T. Fukuda, H. Kimura, H. Nojiri, K. Kaneko, and S. Maekawa, Phys. Rev. Lett. **82**, 4328 (1999).

[8] E. Dagotto, *Nanoscale Phase Separation and Colossal Magnetoresistance* (Springer, New York, 2008).

[9] K. Hirota, N. Kanedo, A. Nishizawa, Y. Endoh, M. C. Martin, and G. Shirane, Physica B **237-238**, 36 (1997).

[10] D. E. Cox, P. G. Radaelli, M. Marezio, and S.-W. Cheong, Phys. Rev. B **57**, 3305 (1998).

[11] P. G. Radaelli, R. M. Ibberson, D. N. Argyriou, H. Casalta, K. H. Andersen, S.-W. Cheong, and J. F. Mitchell, Phys. Rev. B **63**, 172419 (2001).

[12] P. G. Radaelli, R. M. Ibberson, S.-W. Cheong, and J. F. Mitchell, Physica B **276-278**, 551 (2000).

[13] I. G. Deac, J. F. Mitchell, and P. Schiffer, Phys. Rev. B **63**,

172408 (2001).
[14] H. Yoshizawa, H. Kawano, Y. Tomioka, and Y. Tokura, Phys. Rev. B **52**, R13145 (1995).
[15] U. Scotti di Uccio, B. Davidson, R. Di Capua, F. Miletto Granozio, G. Pepe, P. Perna, A. Ruotolo, and M. Salluzzo, J. Alloys Compd. **423**, 228 (2006).
[16] Y. Z. Chen, J. R. Sun, S. Liang, W. M. Lv, B. G. Shen, and W. B. Wu, J. Appl. Phys. **103**, 096105 (2008).
[17] Y. Wakabayashi, D. Bizen, H. Nakao, Y. Murakami, M. Nakamura, Y. Ogimoto, K. Miyano, and H. Sawa, Phys. Rev. Lett. **96**, 017202 (2006).
[18] M. Radovic, N. Lampis, F. Miletto Granozio, P. Perna, Z. Ristic, M. Salluzzo, C. M. Schlepütz, and U. Scotti di Uccio, Appl. Phys. Lett. **94**, 022901 (2009).
[19] C. Barone, A. Galdi, N. Lampis, L. Maritato, F. M. Granozio, S. Pagano, P. Perna, M. Radovic, and U. Scotti di Uccio, Phys. Rev. B **80**, 115128 (2009).
[20] Z. Jirak, S. Krupicka, Z. Simsa, M. Dlouha, and V. Vratislav, J. Magn. Magn. Mater. **53**, 153 (1985).
[21] B. Mercey, P. Lecoeur, M. Hervieu, J. Wolfman, Ch. Simon, H. Murray, and B. Raveau, Chem. Mater. **9**, 1177 (1997).
[22] F. Miletto Granozio and U. Scotti di Uccio, J. Cryst. Growth **174**, 409 (1997); J. Alloys Compd. **251**, 56 (1997).
[23] M. Fujimoto, H. Koyama, Y. Nishi, T. Suzuki, S. Kobayashi, Y. Tamai, and N. Waya, J. Am. Ceram. Soc. **90**, 2205 (2007).
[24] J. W. Seo, E. E. Fullerton, F. Nolting, A. Scholl, J. Fompeyrine, and J.-P. Locquet, J. Phys.: Condens. Matter **20**, 264014 (2008).
[25] A. M. Haghiri-Gosnet, M. Hervieu, Ch. Simon, B. Mercey, and B. Raveau, J. Appl. Phys. **88**, 3545 (2000).
[26] J. M. De Teresa, M. R. Ibarra, J. García, J. Blasco, C. Ritter, P. A. Algarabel, C. Marquina, and A. del Moral, Phys. Rev. Lett. **76**, 3392 (1996).
[27] D. A. Pejaković, J. L. Manson, J. S. Miller, and A. J. Epstein, Phys. Rev. Lett. **85**, 1994 (2000).
[28] A. Geddo Lehmann, C. Sanna, N. Lampis, F. Congiu, G. Concas, L. Maritato, C. Aruta, and A. Y. U. Petrov, Eur. Phys. J. B **55**, 337 (2007).
[29] J. A. Mydosh, *Spin Glasses: An Experimental Introduction* (Taylor & Francis, London, 1993).
[30] J. R. L. de Almeida and D. J. Thouless, J. Phys. A **11**, 983 (1978).
[31] M. Gabay and G. Toulouse, Phys. Rev. Lett. **47**, 201 (1981).
[32] F. Bernardot and C. Rigaux, Phys. Rev. B **56**, 2328 (1997).
[33] G. G. Kenning, D. Chu, and R. Orbach, Phys. Rev. Lett. **66**, 2923 (1991).
[34] D. N. H. Nam, R. Mathieu, P. Nordblad, N. V. Khiem, and N. X. Phuc, Phys. Rev. B **62**, 8989 (2000).
[35] N. X. Phuc, N. Van Khiem, and D. N. H. Nam, J. Magn. Magn. Mater. **242-245**, 754 (2002).
[36] A. Geddo Lehmann, C. Sanna, F. Congiu, G. Concas, and L. Maritato, Phys. Status Solidi B **246**, 1948 (2009).
[37] A. Ruotolo, A. Oropallo, F. Miletto Granozio, G. P. Pepe, P. Perna, and U. Scotti di Uccio, Appl. Phys. Lett. **88**, 252504 (2006).
[38] J. Dho, E. O. Chi, N. M. Hur, K. W. Lee, H. S. Oh, and Y. N. Choi, Solid State Commun. **123**, 441 (2002).
[39] J. Dho, W. S. Kim, E. O. Chi, N. H. Hur, S. H. Park, and H.-C. Ri, Solid State Commun. **125**, 143 (2003).
[40] M. Roy, J. F. Mitchell, A. P. Ramirez, and P. Schiffer, Phys. Rev. B **62**, 13876 (2000).
[41] A. V. Savinkov, D. S. Irisov, B. Z. Malkin, K. R. Safiullin, H. Suzuki, M. S. Tagirov, and D. A. Tayurskii, J. Phys.: Condens. Matter **18**, 6337 (2006).
[42] D. A. Joshi, R. Nagalakshmi, R. Kulkarni, S. K. Dhar, A. Thamizhavel, Physica B **404**, 2988 (2009).
[43] T. Ambrose and C. L. Chien, Phys. Rev. Lett. **76**, 1743 (1996).
[44] A. Biswas, M. Rajeswari, R. C. Srivastava, Y. H. Li, T. Venkatesan, R. L. Greene, and A. J. Millis, Phys. Rev. B **61**, 9665 (2000).
[45] M. Bibes, L. I. Balcells, S. Valencia, J. Fontcuberta, M. Wojcik, E. Jedryka, and S. Nadolski, Phys. Rev. Lett. **87**, 067210 (2001).
[46] S. Estradé, J. Arbiol, F. Peiró, I. C. Infante, F. Sánchez, J. Fontcuberta, F. de la Peña, M. Walls, and C. Colliex, Appl. Phys. Lett. **93**, 112505 (2008).
[47] D. N. Argyriou, U. Ruett, C. P. Adams, J. W. Lynn, and J. F. Mitchell, New J. Phys. **6**, 195 (2004).
[48] A. J. Millis, T. Darling, and A. Migliori, J. Appl. Phys. **83**, 1588 (1998).
[49] J. Kanamori, J. Appl. Phys. **31**, 14S (1960).
[50] H. Y. Hwang, S.-W. Cheong, P. G. Radaelli, M. Marezio, and B. Batlogg, Phys. Rev. Lett. **75**, 914 (1995).
[51] D. P. Kozlenko, V. P. Glazkov, R. A. Sadykov, B. N. Savenko, V. I. Voronin, and I. V. Medvedeva, J. Magn. Magn. Mater. **258-259**, 290 (2003).
[52] J. B. Goodenough, J. Appl. Phys. **81**, 5330 (1997).
[53] F. Rivadulla, M. A. López-Quintela, J. Mira, and J. Rivas, Phys. Rev. B **64**, 052403 (2001).
[54] J.-S. Zhou and J. B. Goodenough, Phys. Rev. B **62**, 3834 (2000).
[55] Y. Zhao, D. J. Weidner, J. B. Parise, and D. E. Cox, Phys. Earth Planet. Inter. **76**, 1 (1993); **76**, 17 (1993).